# Electronic transport properties of topological insulator films and low dimensional superconductors

Ying Xing(邢颖)[1], Yi Sun(孙祎)[1], Meenakshi Singh[2], Yan-Fei Zhao(赵弇斐)[1], Moses H. W. Chan[2], Jian Wang(王健)[1,2]*

[1]*International Center for Quantum Materials, School of Physics, Peking University, Beijing, 100871, China*
[2]*The Center for Nanoscale Science and Department of Physics, The Pennsylvania State University, University Park, Pennsylvania 16802-6300, USA*
E-*mail: jianwangphysics@pku.edu.cn*

In this review, we present a summary of some recent experiments on topological insulators (TIs) and superconducting nanowires and films. Electron-electron interaction (EEI), weak anti-localization (WAL) and anisotropic magneto-resistance (AMR) effect found in TI films by transport measurements are reported. Then, transport properties of superconducting films, bridges and nanowires and proximity effect in non-superconducting nanowires are described. Finally, the interplay between TIs and superconductors (SCs) is also discussed.

## Contents

## 1 Introduction

TI [1] and SC are two important classes of materials being studied in condensed matter physics. While SC continues to hold scientific interest after more than a century of their discovery, TI is a relatively new material that has been the subject of many theoretical and experimental studies for the last decade.

The physics of TIs originates from quantum spin Hall effect [2]. In these materials, a finite energy gap found in the bulk of a TI is crossed by the two gapless surface (3 dimensional (3D)) or edge (2D) states. The two surface state branches have opposite spins cross at a Dirac point and are protected from backscattering by time reversal symmetry. As a result of this unique band structure, TIs behave like insulators in their bulk but show conduction on their surface or edge.

The existence of TI surface states are confirmed by surface probe techniques such as angle resolved photoemission spectroscopy (ARPES) [3; 4] and scanning tunneling microscopy (STM) [5; 6]. However, probing the surface state by transport measurements is quite challenging. This is because the Fermi level of TI samples may not lie in the bulk band gap leading to both surface and bulk states contributing to transport. In spite of this roadblock, much progress has been achieved and ingenious methods devised in the last few years to suppress the bulk conductivity and identify the surface state transport by electrical gating [7] and doping [8]. With the promise of fault proof quantum computing and efficient thermo-electrics amongst others, TIs continue to evolve rapidly and reveal new and exciting physics.

In contrast to TI, SC has a century long revered research history. For over 7 decades of those 100 years, studies on SCs were largely focused on bulk properties. As such, the behavior of bulk SCs is reasonably well understood theoretically and experimentally. Electron transport measurements have been the workhorse for studying superconductivity and have played a decisive role in the search for new superconductors. With advances in synthesis techniques, spatially confined superconducting elements came into existence. The properties of these low dimensional superconductors can be quite different from bulk superconductors since thermal and quantum fluctuations of the order parameter begin to play an increasingly important role. Moreover, extremely large surface-to-volume ratio in these systems also makes surface effects very important. As such, low-dimensional SCs open new avenues for exploration of novel quantum phenomena as well as potential applications. Both TIs and low dimensional SCs are fertile grounds for scientific exploration. Elusive physics like that of the Majorana fermion is expected to exist at the interface of the two. As such, studying these systems individually and in conjunction is a promising venture.

In this review, we introduce our results on transport properties of TI films in section 2, including EEI, WAL and AMR. In section 3, we summarize the results of transport experiments on nanostructured conventional SCs, such as superconducting bridges and nanowires. Section 4 contains experimental results detailing the interaction between TI films (nanobelts) and superconducting electrodes. Finally, a brief overview of these results and future prospects is given in Section 5.

## 2 Transport properties of TI films

TIs represent a unique phase of quantum matter with an insulating bulk gap and gapless edges (or surface) states. These states are possible due to the combination of spin-orbit interactions and time-reversal symmetry. The properties of the exotic electronic states of TIs have been uncovered in several transport measurements [9; 10; 11; 12]. Since the first predicted topological insulator, 2D HgTe [13], many new TI materials have been identified and studied. Among these, $Bi_2Se_3$ has a simple band structure with a single Dirac cone on the surface and a relatively large non-trivial bulk band gap of 0.3 eV [14]. These properties make $Bi_2Se_3$ ideal for the study of interesting topological phenomena. Here we focus on the EEI, WAL and AMR properties in $Bi_2Se_3$ films grown by MBE.

2.1 EEI and WAL

WAL is always expected in systems with either strong spin-orbit scattering or coupling. The spin of the carrier rotates as it goes around a self-intersecting path, and the direction of this rotation is opposite for the two directions about the loop. Because of this, the two paths any loop interfere destructively which leads to a lower net resistivity. The WAL is suppressed by applying a magnetic field, thus giving rise to a negative magneto-conductivity. Because of the robustness of the surface state, the WAL is protected against the strength of disorder and nonmagnetic impurity. Magnetic impurities doping breaks the time-reversal symmetry and destroys the TI state, causing a crossover from the WAL to regular weak localization (WL) [15]. Since the transport properties of 3D TI were firstly investigated by N.P.Ong and collaborators [9], WAL theory is widely studied in this field [15; 16; 17; 18; 19; 20]. Among these studies, our group first proved the importance of EEI except for the WAL effect. In our experiment, the thickness of the topological insulator $Bi_2Se_3$ films used is smaller than the inelastic mean free path, which makes the two dimensional insofar as the WAL correction to conductivity is concerned. The details of the experiment and the theoretical analysis are as follows.

In this work, two single crystal thin films of $Bi_2Se_3$, grown by Molecular Beam Epitaxy (MBE), both with and without Pb doping were considered. 45 nm $Bi_2Se_3$ films with 0.37% Pb doping were grown on bare insulating 6H-SiC (0001) substrates. The resistivity of the substrate was as large as $1 \times 10^6$ $\Omega\cdot$cm. After doping, ARPES results [21] show that the Fermi level of the $Bi_{2-x}Pb_xSe_3$ film was inside the bulk energy gap. Thus, the bulk conductivity should be suppressed and the surface conductance should come into evidence. Before the samples were taken out of the MBE chamber for *ex situ* measurements, a 30 nm thick amorphous Se layer was deposited on the films as a protective layer. Transport measurements were carried on in a Physical Property Measurement System (PPMS).

The conductance as a function of temperature and magnetic field for the $Bi_2Se_3$ and $Bi_{2-x}Pb_xSe_3$ films is shown in Fig. 1. We first attempt to analyze these results using the standard results of WL theory. We know that in perpendicular magnetic field:

$$\Delta\sigma_{WL}(H_\perp) - \Delta\sigma_{WL}(0) = \alpha \frac{e^2}{2\pi^2\hbar} \quad (1)[22]$$
$$\times \left[ \Psi\left(\frac{1}{2} + \frac{\hbar c}{4e\ell_\phi^2 H_\perp}\right) - \ln\left(\frac{\hbar c}{4e\ell_\phi^2 H_\perp}\right) \right]$$

In parallel magnetic field:

$$\Delta\sigma_{WL}(H_\parallel) - \Delta\sigma_{WL}(0) = \quad (2)[23]$$
$$-\frac{e^2}{4\pi^2}\left\{ \ln\left|1 + 2\left(\frac{x+z}{x-y}\right)\frac{\tau_\phi}{\tau}\right| \right.$$
$$\left. -\frac{1}{\sqrt{1-\gamma}} \ln \frac{\frac{\tau_\phi}{\tau} + \left(\frac{x+z}{x-y}\right)(1+\sqrt{1-\gamma})}{\frac{\tau_\phi}{\tau} + \left(\frac{x+z}{x-y}\right)(1-\sqrt{1-\gamma})} \right\}$$

Where, $x = \frac{1}{\tau_0}$, $y = \frac{1}{\tau_{s0,\perp}}$, $z = \frac{1}{\tau_{s0,//}}$

$$\gamma = \left(\frac{g\mu_B B\tau}{\hbar}\left(\frac{x-y}{y+z}\right)\right)^2$$

Where, α is a constant depending on the relative strengths of the spin-orbit and spin-flip (magnetic) scattering, ψ is the digamma function, $\ell_\Phi$ is the inelastic scattering length, g is the Zeeman g-factor, $\tau_0$ refers to the scattering time from disorder, $\tau_{so,i}$ refers to the spin-orbit scattering in the directions perpendicular and parallel to the film. Explicit comparison of the experimental results to the WL theory can be seen in Fig. 1. When the magnetic field is perpendicular to the film, The fitting lines in Figs. 1(c) and 1(d) show the WL theory (Equation (1)) for α = 1, as implied by the positive slopes of the data in Figs. 1(a) and 1(b). Figures 1(e) and 1(f) show the magneto-conductance in parallel fields ($H_\parallel$ denotes a field parallel to the film and perpendicular to the excitation current, while $H_\parallel$ denotes a field parallel to both the film and the current). Lines are best fits to the theory of Maekawa and Fukuyama.

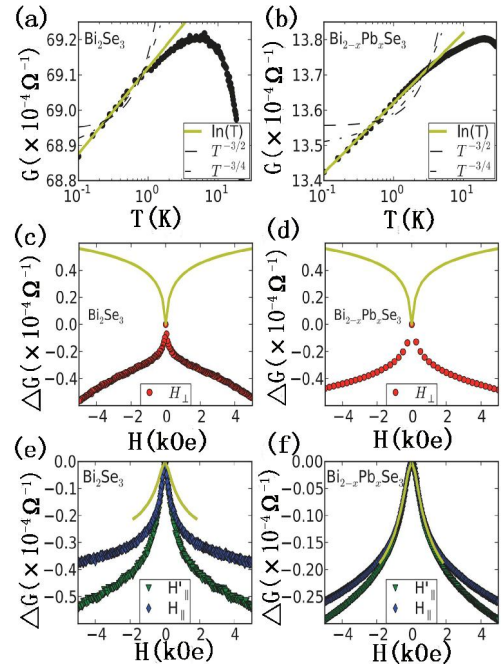

**Fig. 1** WL theory **(a)** and **(b)**, Temperature dependence of the conductance. Lines are fits to the WL theory in 2D (solid) and 3D (dashed and dash-dotted). The temperature dependence of both films suggests that our films are in the regime of weak spin-orbit scattering. However, in this regime, the theory predicts a positive magnetoconductance in perpendicular field. **(c)** and **(d)**, Magnetic field dependence of ΔG in a perpendicular magnetic field. ΔG = ΔG(H)−ΔG(0). The theory predicts a localization effect; however, we observe antilocalization. **(e)** and **(f)**, Magnetoconductance in parallel fields ($H_\parallel$ denotes a field parallel to the film and perpendicular to the excitation current, while $H_\parallel'$ denotes a field parallel to both the film and the current). This figure is from Ref. [21].

Since WL theroy does not fit the experimental data

well, EEI has to be taken into account:

$$\Delta\sigma_{EEI}(T) = \frac{e^2}{4\pi^2\hbar}(2 - \frac{3}{2})\tilde{F}_\sigma \ln(\frac{T}{T_0}) \quad (3)[24]$$

Where, $T_0$ is a reference temperature from which one measures the deviation $\Delta\sigma$, $\tilde{F}_\sigma$ is a function of the average of the static screened coulomb interaction over the Fermi surface. This expression neglects the spin-orbit scattering. Figure 2 shows a comparison of the same data as Fig. 1 with the EEI theory. Figures 2(a) and 2(b) show the transport properties in low temperature and low magnetic field. The left columns are for the undoped $Bi_2Se_3$ film, while the doped $Bi_{2-x}Pb_xSe_3$ data are on the right. Conductance *vs.* T for zero magnetic field (black circles) and for $H_{//}$ = 20 kOe (red triangles), with fits to 2D theories (solid lines), and 3D EEI theory (dash-dotted lines) are shown in Figs. 2(a) and 2(b). Figures 2(c)-2(f) show the magnetoconductance $\triangle G = G(H) - G(0)$ for fields perpendicular and parallel to the film. Solid lines are the result of EEI theory. Although it does not capture the sharp peaks at zero magnetic field, the EEI expressions correctly reproduce the signs of the T and H dependences. This indicates the importance of the EEI interaction in transport properties.

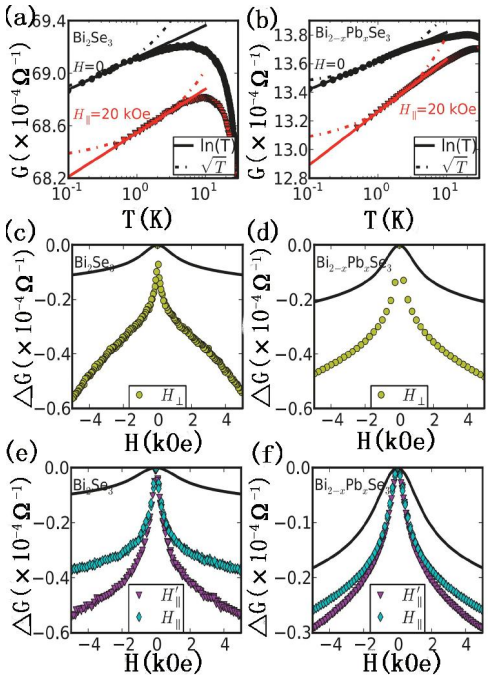

**Fig. 2** EEI theory **(a)** and **(b)** Temperature dependence of the conductance. Lines are fits to the 2D(solid) and 3D(dashed) theory of Lee and Ramakrishnan. **(c)** and **(d)** Magnetic field dependence of ΔG in a perpendicular magnetic field. **(e)** and **(f)** Magnetic field dependence of ΔG in parallel field. Solid lines in (c)–(f) are fits to the EEI theory using no additional fitting parameters. This figure is from Ref. [21].

Finally, attempts are made to fit the experimentally measured data to a combination of WAL and EEI. The WAL term also includes a factor α akin to that seen in the weak localization term (2). This constant factor depends on the relative strengths of the spin-orbit and spin-flip (magnetic) scattering. For the fits shown in Fig. 3, conductance *vs.* temperature for zero magnetic field (black circles) and for $H_{//}$ = 20 kOe (red triangles), with fits to 2D theories (solid lines), and 3D EEI theory (dash-dotted lines) are shown in Figs. 3(a) and 3(b). Figrues 3(c)-3(f) show the magnetoconductance $\triangle G = G(H) - G(0)$ for fields perpendicular and parallel to the film. A very satisfactory fit is obtained by using the following form of conductivity:

$$\triangle\sigma(T, H) = \triangle\sigma_{EEI}(T, H) + \triangle\sigma_{WAL}(T, H) \quad (4)$$

Where, α = 1 represents the limit of weak spin orbit and magnetic scattering; α = –1/2 represents the limit of strong spin-orbit scattering and weak magnetic scattering. We find α = –0.31(–0.35) for $Bi_2Se_3$ ($Bi_{2-x}Pb_xSe_3$) which is close to strong spin-orbit interaction. 2D WAL model is more relevant than 3D, because the inelastic scattering length in the two samples is much larger than the samples thickness.

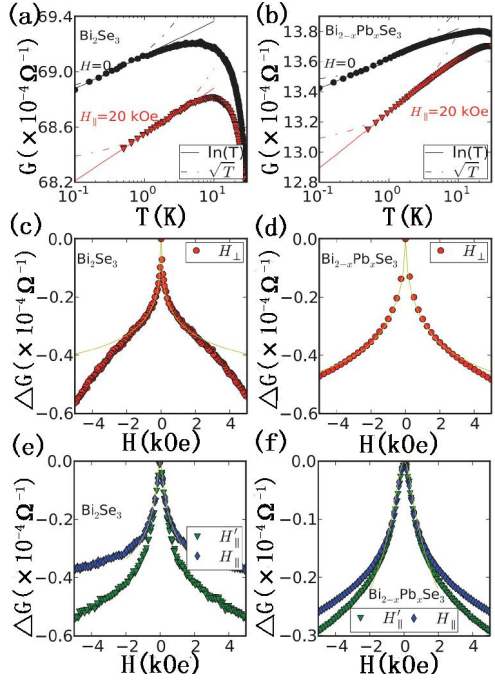

**Fig. 3** Combined anti-weak localization and EEI interaction **(a)** and **(b)** Temperature dependence of the conductance with fits to 2D theories (solid lines), and 3D electron-electron interaction (EEI) theory (dash-dotted lines). **(c)** and **(d)** Magnetic field dependence of ΔG in a perpendicular magnetic field. Solid lines are the result of a combined WL and EEI theory. **(e)** and **(f)** Magnetic field dependence of ΔG in parallel field. The solid lines are fits to the combined WL and EEI theory. This figure is from Ref. [21].

These results clearly demonstrate that it is crucial to include EEI for a comprehensive understanding of diffusive transport in TIs. While both the ordinary bulk and the topological surface states presumably participate in transport, this analysis does not allow a clear separation of the two contributions. Similar WAL combined with EEI is also observed in $Bi_2Te_3$ and $Sb_2Te_3$ film on GaAs(111) substrates, which may pave a potential route to fabricate topological p-n junctions [25].

## 2.2 Anomalous AMR

AMR effect is a property of a material in which the magneto-resistance depends on the angle between the electric current and the magnetic field. In our high quality MBE-grown $Bi_2Se_3$ TI thin films, measurements in a magnetic field in the plane of the substrate, parallel and perpendicular to the bias current show anomalous and opposite MR [21].

Figure 4 shows the magneto-resistance properties under an in-plane magnetic field (from 80 kOe to -80 kOe). When the field is perpendicular to the current and crystal axis, the MR is positive as shown in Figs. 4(a) and 4(c). At low temperatures, an MR dip appears around zero magnetic field due to WAL effects. The MR dip at small field decreases with increasing temperature and disappears around 20 K. The MR changes to negative when the magnetic field is parallel to the current as shown in Figs. 4(b) and 4(d). The positive MR dip at low field that exists at low temperature also disappears above 20 K. A possible mechanism may be that the Lorentz force deflects the surface electrons, leading to a positive MR from the classical galvanomagnetic effect. The effect from the angle between the spin polarization of surface current and magnetic field direction appears to "overcome" the positive MR effect resulting in negative MR behavior.

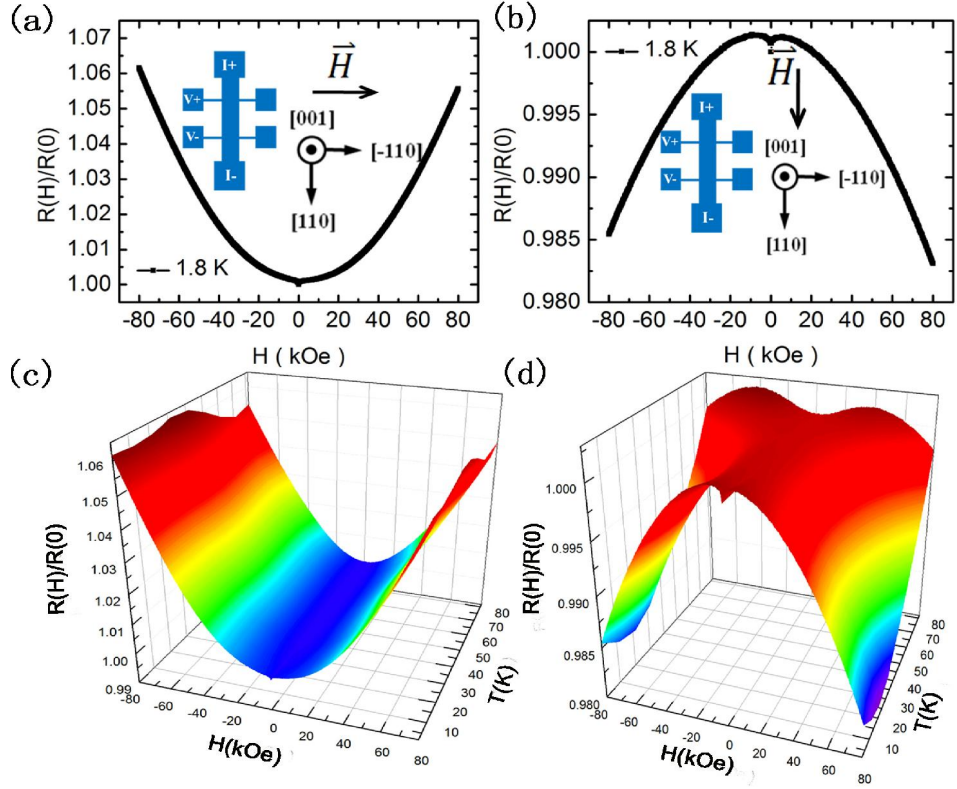

**Fig. 4** In plane magneto-resistance of 200 nm thick $Bi_2Se_3$ film. **(a)** The magnetic field is perpendicular to the current direction. **(b)** The magnetic field is parallel to the current direction. **(c)** Three-dimensional image of the magneto-resistance at different temperatures when the field is perpendicular to the current. **(d)** Three-dimensional image of the magneto-resistance at different temperatures when the field is parallel to the current. This figure is from Ref. [26].

---

# 3 Experimental progresses in low dimensional SCs

Nanoscale SCs grown on semiconductor substrates is one of the most attractive research fields since the derived SC-based electronics have been shown to be promising for future processing and storage technologies. By utilizing electrochemical deposition, MBE, focused ion beam etching and depositing system (FIB) and PPMS, we investigated the transport properties of conventional superconducting films, bridges, nanowires and the proximity effect in non-superconducting nanowires.

3.1 Pb nanostructures

*3.1.1 Unusual resistance and magneto-resistance*

The resistance vs. temperature (*R-T*) behavior (Fig. 5(b)) of smooth Pb film (Fig. 5(a)) shows an abrupt drop to zero resistance at a well defined transition temperature. On the other hand for a fractal-like film (Fig. 5(c)), the resistance shows a comparatively small drop and then an increase with decreasing temperature Fig. 5(d)). The fractal-like morphology Pb film was formed as a result of exposing the flat Pb film at room temperature in atmosphere for 48 h. As expected, the resistance of the fractal film is much higher than that of the smooth film. With decreasing temperature, the resistance of the fractal film drops rapidly and reaches a minimum at 5.4 K, but increases as the temperature continues decreasing. The superconductivity onset $T_C$ of the smooth film (Fig. 5(b)) is 6.1 K, lower than onset $T_C$ (7.0 K) of the fractal film (Fig. 5(d)).

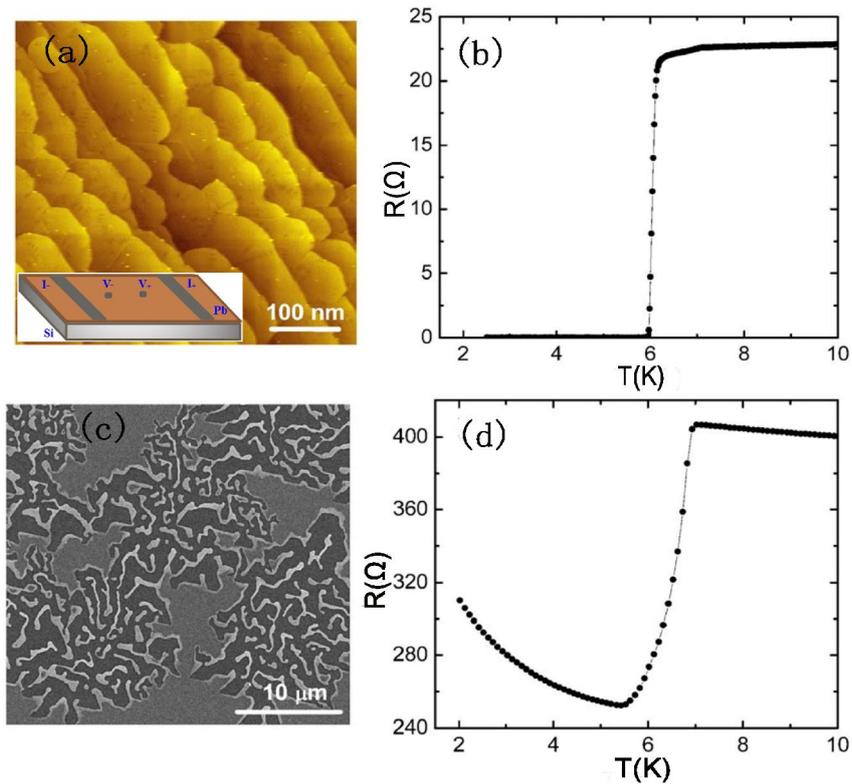

**Fig. 5** (a) STM image (500×500 nm$^2$) of the 23 atomic monolayers atomically flat Pb thin film, the inset is the schematic graph for the transport measurement. (b) $R$ vs. $T$ curve measured from the Pb film in (a). (c) Scanning electron micrograph (SEM) of the fractal-like Pb film, the dark regions are the Si substrate and the gray and white ones are the Pb films. (d) $R$ vs. $T$ curve of the fractal-like Pb film shown in (c). This figure is from Ref. [27].

We cut a uniform flat film into two parts with a 2 μm wide gap etched by FIB (Fig. 6(a) shows an SEM of the film thus cut). The resistance-magnetic field ($R$-$H$) (Fig. 6 (b)) curve exhibits anomalous enhancement at zero magnetic field. The magneto-resistance peak at zero field decreases rapidly with increasing temperature. This unusual magneto-resistance effect in superconductor–semiconductor heterojunctions may be utilized in developing a magnetic field controlled "on–off" device or a high-sensitivity field sensor.

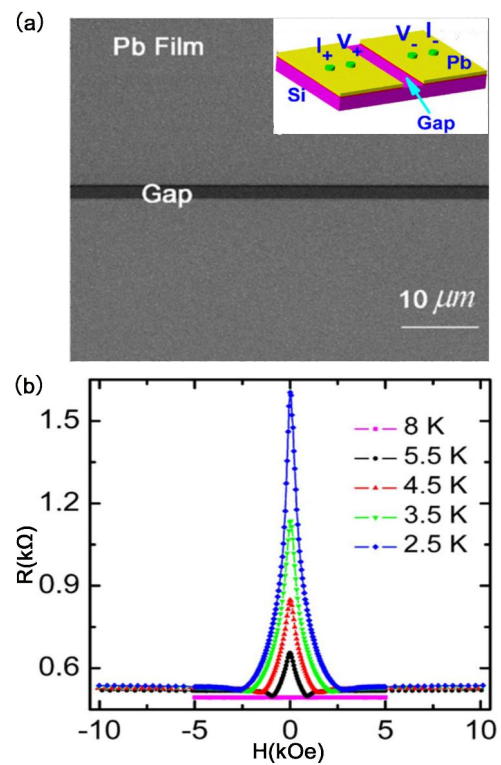

**Fig. 6 (a)** A SEM of 26 atomic monolayers Pb film after a 2 μm wide gap (the dark region) was fabricated, the inset is the schematic graph for the transport measurement. **(b)** Magneto-resistance of the heterojunctions with a magnetic field perpendicular to the film at different temperatures. This figure is from Ref. [28].

*3.1.2 Magneto-resistance oscillations*

The smooth Pb films were etched using FIB and crystalline Pb nanobelts were thus fabricated (Fig. 7 (a)). Anomalous magneto-resistance oscillations as well as enhanced superconductivity were observed (Fig. 7(b)). Compared with crystalline Pb film ($T_C$ = 6.3 K) grown by MBE, the $R$ vs. $T$ curve of Pb nanobelt (285 nm wide) shows a broader superconducting transition and a significantly higher onset $T_C$ (6.9 K) although both of them are lower than the $T_C$ of the bulk Pb (7.2 K).

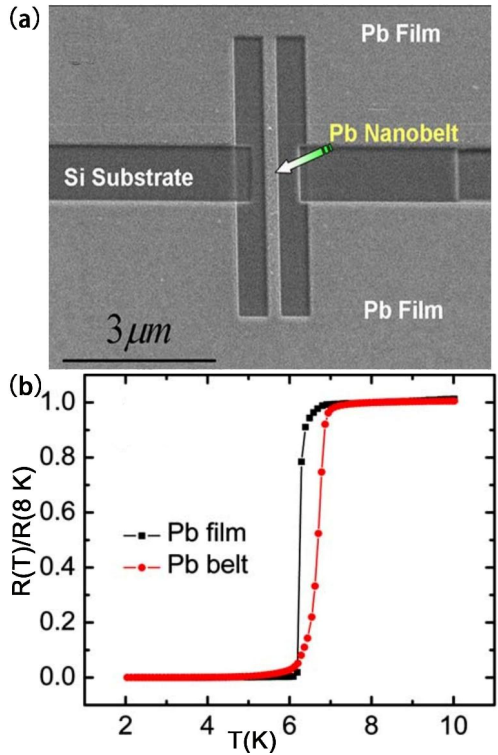

**Fig. 7 (a)** SEM of the Pb nanobelt made of the Pb film. The Pb nanobelt is 28 atomic monolayers thick, 285 nm wide and 10 μm long. The dark region on the two sides of the Pb nanobelt is exposed Si surface, which isolates two blocks of the Pb film. **(b)** Resistance as a function of temperature measured from the Pb film and the Pb nanobelt, respectively. This figure is from Ref. [29].

Magneto-resistance oscillations are observed in Pb nanobelts for temperatures below $T_C$ (see Fig. 8) [30]. The physical mechanism behind these oscillations is not fully understood yet. However, one possible explanation could be that there are mesoscopic superconducting ring-like structures in the Pb nanobelt due to inhomogeneous superconductivity. The existence of these superconducting rings may give rise to Little-Parks [31] like periodic magneto-resistance oscillations in the superconducting regime, such as Little–Parks-like oscillations.

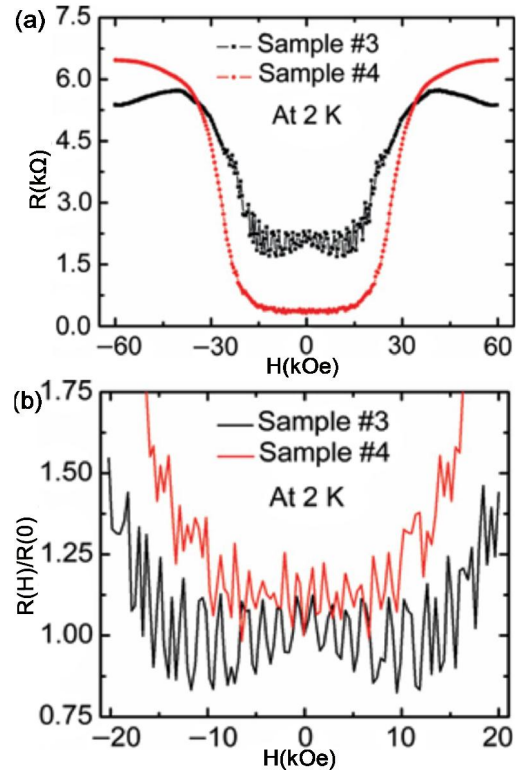

**Fig. 8 (a)** Magneto-resistance of two different samples with a magnetic field applied perpendicular to the film at 2 K for comparison. (#3：2 μm×350 nm×29 Molecule Layers；#4：10 μm×284 nm×28 Molecule Layers) **(b)** Close-up view of (a) in the low magnetic field regime at 2 K for clarity. The vertical scale is normalized to the resistance at zero field. This figure is from Ref. [30].

3.2 Superconductivity in crystalline Bi nanowire

Although bulk Bi is a semimetal, we have also found clear evidence of superconductivity in cystalline Bi nanowires of 72 nm diameter [32].

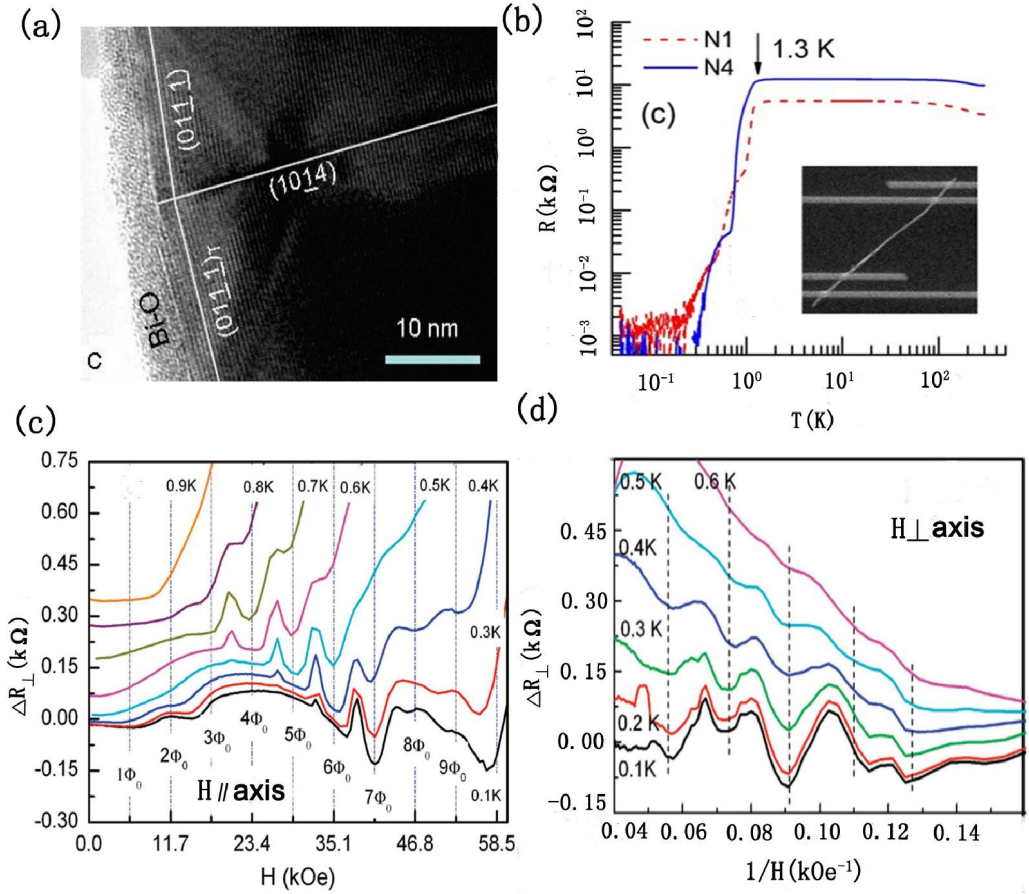

**Fig. 9 (a)** High-resolution transmission electron microscope (HRTEM) image near one of these stripes, indicating that the stripes are actually (1014) twinning boundary, which is perpendicular to the growth direction. **(b)** $R$ vs. $T$ curves of sample, measured under different perpendicular magnetic fields. Four Pt leads deposited by FIB technique is shown in the inset of (b). **(c)** $\Delta R_{\parallel}$-$H$ and **(d)** $\Delta R_{\perp}$-$H$ plots in parallel and perpendicular magnetic field. This figure is from Ref. [32].

The Bi nanowires were fabricated using template based electrodeposition [33]. An oxide (BiO) layer of approximately 3.7±0.5 nm (visible on the surface (see Fig. 9(a))) probably formed due to environmental oxidation after the nanowires were released from the membrane. Small angle twinning boundaries (Fig. 9(a)) with a twinning plane of (10$\underline{1}$4) appear perpendicular to the growth direction of the wire. From the $R$ vs. $T$ curve, the nanowire displays a semiconductor like behavior. A superconducting transition and a small resistance shoulder are found at 1.3 K and 0.67 K, respectively (Fig. 9(b)). Under a parallel magnetic field, periodic Little-Parks-like resistance oscillations are clearly seen in the superconducting state (Fig. 9(c)). The dashed lines indicate the positions of fluxoid quantization as predicted by $H(\pi d^2/4) = n\Phi_0$ ($\Phi_0 = h/2e = 2.07 \times 10^{-7}$ Gcm$^2$, here e is the electron charge and $h$ is Plank's constant) with d = 67 nm. By applying a perpendicular field, quasi-periodic oscillations of residual resistance (below $T_C$) with 1/H are clearly seen (Fig. 9(d)). Dashed lines, separated by $\Delta(1/H) = 0.0176$ kOe$^{-1}$, show good correlation with the minima of resistance oscillations as predicted by the SdH (Shubnikov deHaas) effect [34]. We conclude that there is a coexistence of superconducting state (as required for Little parks) and metallic states (as required for SdH) in the surface shell of the Bi nanowire below $T_C$.

3.3 Proximity effect in nanowires

When a SC is brought in contact to a normal metal,

the normal metal acquires some superconducting properties and the superconductivity of the SC is suppressed or weakened. This phenomenon is known as the superconducting proximity effect. The superconducting proximity effect has been studied since the 1960s [35]. The physical mechanism behind the proximity effect is that the Cooper pairs from the SC penetrate into the normal metal over a distance scale for some distance called the "normal metal coherence length": $\xi_N = (\hbar D_N / 2\pi k_B T)^{1/2}$, where $D_N$ is the diffusion constant of the normal metal. This effect is usually unobvious in bulk samples and becomes more pronounced in systems with reduced dimensions. New insights continue to emerge from measurements such as proximity effect in SC-N (normal metal) and SC-FM (ferromagnet) nanowires. Some recent experimental results are recounted below.

### 3.3.1 Superconducting nanowire – normal metallic electrode

Superconducting nanowires are contacted with normal metallic electrodes in order to study their transport properties. According to the proximity effect, around the contact region, part of normal elecrodes become superconducting and the superconductivity of superconducting nanowire is weakened by the normal electrodes.

In our experiment, single crystalline Pb nanowires of different diameters are fabricated by elecro-chemical deposition in AAO membrane with nanopores [36]. Four normal Pt electrodes are deposited on one individual Pb nanowire by FIB. In the superconducting transition region, resistance-temperature (*R-T*), resistance-magnetic field (*R-H*) and resistance-current (*R-I*) scans all show a series of resistance steps with increasing temperature, magnetic field and excitation current respectively. The *R-H* curves at different temperatures for a 55 nm and 70 nm Pb nanowire are shown in Fig. 10 (a) and (b) respectively. The resistance of the nanowire does not reach zero even at 2 K. Larger residual resistance is found in 55 nm Pb nanowire than 70 nm Pb nanowire.

We attribute these phenomena to the inhomogeneity of the wire (origining from FIB-assisted electrodes deposition) and the proximity effect due to the normal metal (Pt) electrodes, which weakens the superconductivity of the Pb nanowires.

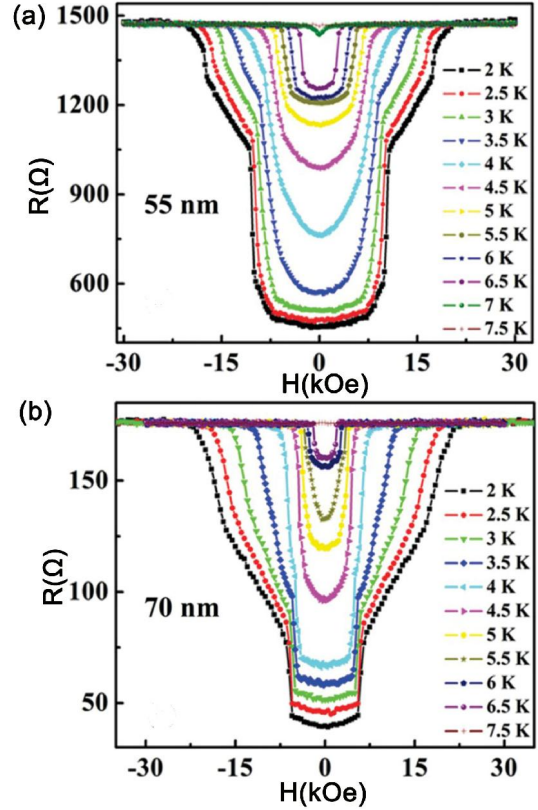

**Fig. 10**. Resistance vs magnetic field of 55 nm **(a)** and 70 nm **(b)** Pb nanowires contacted by FIB fabricated Pt electrodes at different temperatures. This figure is from Ref. [36].

### 3.3.2 Superconducting nanowire – superconducting electrode

As discussed in the previous section, the presne of a bulk SC induces superconductivity in a normal metallic nanowire as predicted by the well understood superconducting proximity effect. In a similar geometry, if a superconducting nanowire was used instead of a normal nanowire, the bulk SC was found to weaken or even suppress its superconductivity. This counterintuitive effect was discovered in and limited to 1D nanowires and was named the anti-proximity effect (APE)[37; 38; 39; 40; 41].

The APE has been studied in single-crystalline Al[39] and Zn[37; 38] nanowires and also granular Al and Zn[40; 41] nanowires. Al and Zn are used because their bulk superconducting coherence length (ξ) is large making the 1D regime (diameter of nanowire < ξ) easily accessible. One of the several experiments is described here. Single-crystalline Al nanowires 70 nm and 200 nm in diameter were synthesized using template based electrochemical deposition[42], into the pores of an anodized aluminum oxide membrane. The nanowires were found to be good quality single-crystal with ~ 5 nm oxide layer on the surface (Fig. 11 (a)). Figure 11(b) shows the normalized voltage vs. applied current for two 70 nm diameter, 2.5 μm long single Al nanowires at 0.1 K. One of the nanowires is measured using normal Pt electrodes and the other is measured using superconducting W electrodes. The inset shows a scanning electron micrograph of the Al nanowire contacted with the Pt electrodes and represents the measurement geometry. The critical current ($I_C$) of the nanowire with superconducting W electrode is much lower than the $I_C$ for the nanowire with the normal Pt electrode. A lower $I_C$ indicates a weakened superconductivity implying that the superocnducting electrode is weakening the superconductivity of the superconducting nanowire. The same effect was not seen in 200 nm diameter Al nanowires similarly measured indicating that this effect is limited to small diameter (1D) nanowires.

One possible explanation of the APE is that it is an experimental manifestation of the Caldeira-Leggett model[43; 44] and can be used as a platform to study macroscopic quantum phenomena like quantum phase slips.

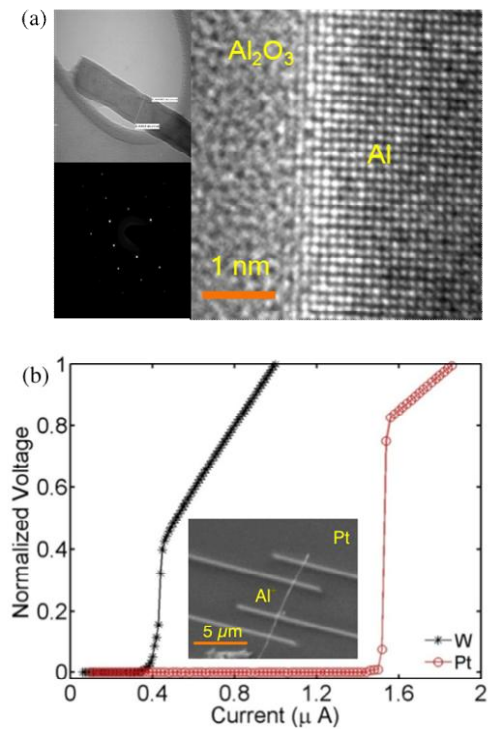

**Fig. 11**. (a) High resolution TEM image of an Al nanowire is shown in the right panel. The oxidation layer and the crystalline nature of the nanowire can be seen. The top left panel shows a less magnified view of the same nanowire and the bottom left panel shows the electron diffraction pattern. (b) shows normalized voltage vs. applied current for two 70 nm diameter, 2.5 μm long single ANWs at 0.1 K. One of the nanowires is measured using normal Pt electrodes and the other is measured using superconducting W electrodes. The inset shows a scanning electron micrograph of the ANW contacted with the Pt electrodes. This figure is taken from Ref. [39].

*3.3.3 Normal metallic nanowire – superconducting electrode*

To study proximity effect induced superconductivity in normal metals, single crystalline gold (Au) nanowires 70 nm in diameter were fabricated using template based electrodeposition in the pores of track-etched polycarbonate membrane [45]. Four superconducting W electrodes were deposited onto one individual Au nanowire by FIB technique. These W strips electrodes are amorphous and composed of tungsten, carbon and gallium[46]. A superconducting transition around 5 K was observed in these W strip by a standard four-probe measurement[46]. Additionally, the W strip shows votex glass to liquid transition-like behavior revealed by voltage vs. currrent measurements[46]. A schematic of the measurement geometry is shown in the inset of Fig. 12(a). The length (1 μm, 1.2 μm, 1.9 μm) of Au nanowire is defined as the distance between the inner edges of the two voltage electrodes.

Four probe transport measurements were made in a PPMS cryostat. The experimental results show all three wires exhibit superconductivity with an onset $T_c$ near 4.5 K (Fig. 12 (a)). Zero resistance was found below 4.05 K for the Au nanowire of 1μm in length. The 1.9 μm long Au nanowire failed to reach zero resistance even at 2 K (Fig. 12 (a)). Interestingly, the resistance drop occured in two steps for the 1.2 μm wire. In the first step between 4.5 and 4.14 K, the resistance is reduced down to 16% of its normal state value, this drop is followed by a more gradual decrease to zero resistance at 3.43 K. The $R$ vs. $H$ curves for three samples at different temperatures are shown in Figs. 12 (b), (c) and (d). The 1 μm long Au nanowire shows typical superconducting behavior while the two other wires show novel mini magneto-resistance valleys near zero magnetic field at low temperature.

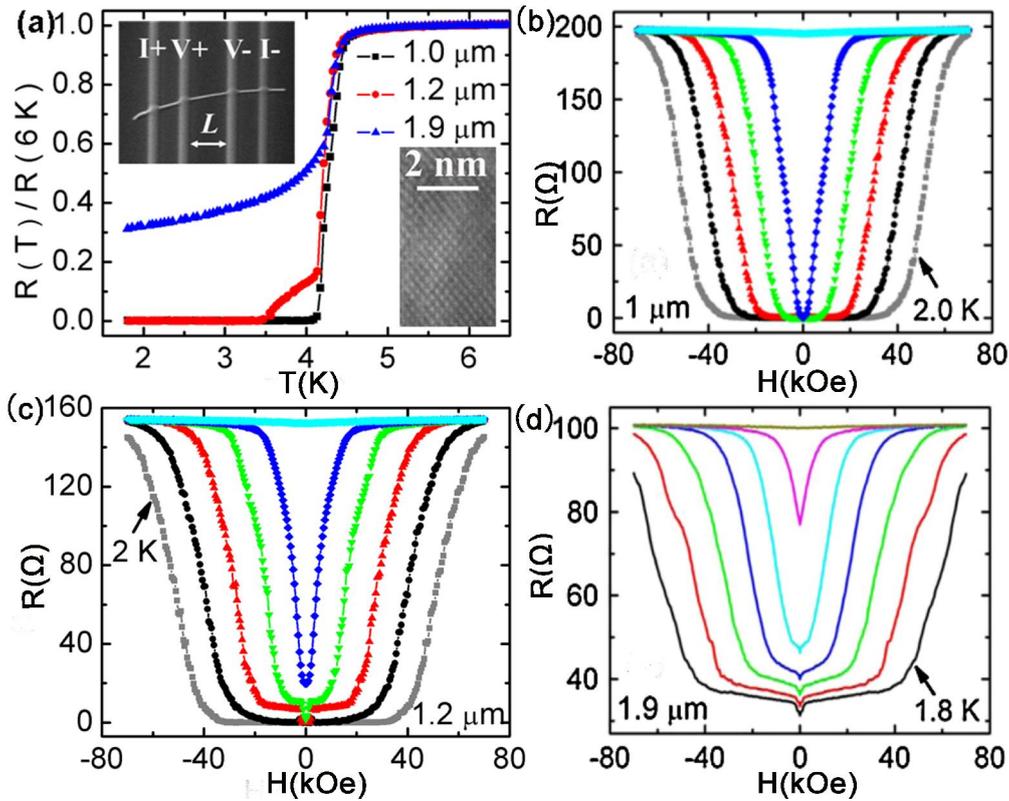

**Fig. 12 (a)** $R$ vs. $T$ curves for individual 70 nm diameter crystalline Au nanowires with lengths (L) of 1μm, 1.2μm and 1.9μm. The vertical scale is normalized to the resistance at $T$ = 6 K. The top left inset is a SEM of an individual 70 nm Au nanowire. The bottom right inset shows HRTEM of a free-standing crystalline Au nanowire showing atomic structure. **(b)** and **(c)** Magnetoresistance of the 1 μm, 1.2 μm Au nanowire, from bottom to top, at 2.0 K (gray), 2.5 K (black), 3.0 K (red), 3.5 K (green), 4.0 K (blue), and 5.5 K (cyan). The magnetic field was applied perpendicular to the axis of the nanowire. **(d)** Magnetoresistance of the 1.9 μm Au nanowire, from bottom to top, at 1.8 K (black), 2.3 K (red), 2.8 K (green), 3.3 K (blue), 3.8 K (cyan), 4.3 K (magenta), and 5.8 K (dark yellow). This figure is from Ref. [47].

We provide a qualitative explanation for these observations [47]. The proximity effect induced superconducting gap as a function of distance 'x' from the electrode in zero magnetic field can be written as

$$\Delta(x) = \Delta_a \cosh(x/\xi_N) / \cosh(a/\xi_N) \quad (4)$$

Where, 2a is the length of the nanowire, $\xi_N$ is coherence length characterizing the decay of the induced superconductivity in Au, $\Delta_a$ is the superconducting gap at the boundary. This can be used to calculate $\Delta_b$, the superconducting gap in the middle of the Au nanowire: $\Delta_b = \Delta(x=0) = \Delta_a \cosh^{-1}(a/\zeta_N)$. For the short (1 μm) nanowire, $\Delta_a$ is on the same order of $\Delta_b$, so the superconductivity is destroyed simultaneously by increasing temperature or magnetic field. For very long nanowire (1.9 μm), the gap in the middle is zero and the middle of the nanowire away from the superconducting electrodes is always normal. Residual resistance exists even at very low temperature. For the medium length nanowire (1.2 μm), with increasing temperature, only the middle of nanowire becomes normal above the first critical temperature (3.43 K) corresponding to the gap closing somewhere in the middle of the nanowire. The whole nanowire becomes normal at a higher temperature (4.14 K). We can see these two resistance transitions in Fig. 12(a).

With an increasing magnetic field, the differential magneto-resistance $dR/dB$ of the 1μm and, 1.2μm nanowire beyond the critical breakdown magnetic field $H_b$ shows uniform oscillations with a period of $\Phi_0/(2\pi r^2)$ ($\Phi_0 = h/2e$ is the superconducting flux quantum, r = 35 nm is the radius of the nanowire) below $T_C$ (Fig. 13 (a)). In the 1.9 μm Au nanowire, there is no well defined periodicity in the oscillations. And in the measurement of a W strip with four W electrodes this phenomenon was not found. Following the same model for the superconducting gap in the nanowire (equation (4)), this phenomenon can also be explained (Fig. 13 (b)). These oscillations may arise from the sequential generation and moving of vortices. Because the gap is different at different points along the nanowire, vortices are introduced into different length segments at different applied fields. With increasing magnetic field, the generated vortices move continuously across the wire and therefore the resistance of the Au wires increases in a stepwise fashion.

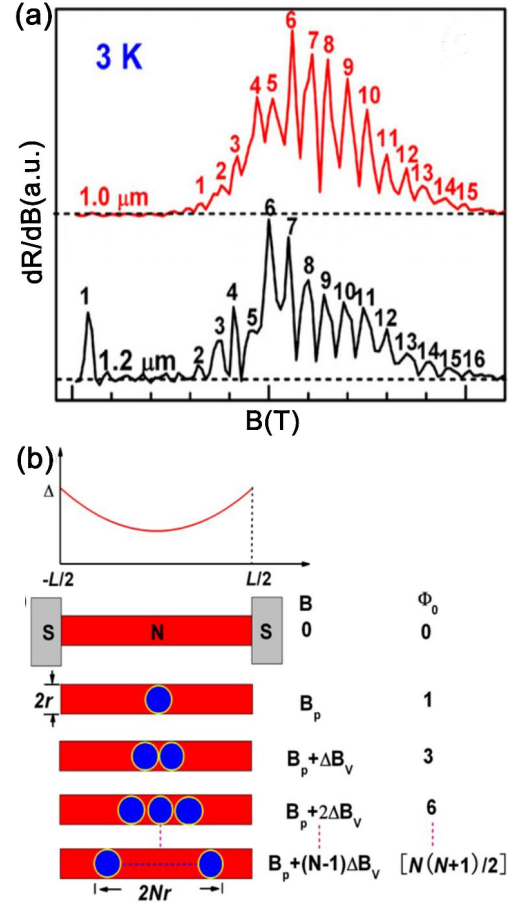

**Fig. 13 (a)** The d$R$/d$B$ curves of the 1.0 and 1.2 μm Au nanowires measured at $T$ = 3 K. The dashed lines represent d$R$/d|$B$| = 0 for the two wires. There are small oscillations of the value of d$R$/d|$B$| around zero, possibly due to the limited resolution in the $R$ and $B$ readings in low magnetic field. We picked the clearly resolved peak at about 1.65 T as the first peak of the 1.0 μm wire. The peak at about 1.6 T of the 1.2 μm Au nanowire was numbered as the N = 2 peak. Except for the first peak of the 1.2 μm nanowire, the differential magnetoresistance shows uniform oscillations with $B$ of 0.25 T. **(b)** The superconducting gap as a function of the position induced by the proximity effect in the nanowire. Schematic view of the SC–NW–SC structure and the vortices induced by the applied field. Below the critical field Bp, all flux is expelled from the nanowire. At Bp, one vortex carrying one flux quantum $\Phi_0$ enters the nanowire. Above this critical field, additional vortices are induced one at a time with magnetic field widths of $\Delta B_V = \Phi_0/2\pi r^2$. This figure is from Ref. [48].

### 3.3.4 Ferromagnetic nanowire – superconducting electrodes

The proximity effect spatial range of SC/N interface can be as long as 1 μm. In a FM/SC interface however, the contradicting spin orders between the FM and the singlet SC are expected to greatly reduce this range. In a FM/s-wave SC interface, superconductivity is expected to decay rapidly (in a few nanometres) inside the FM [49]. When a conventional spin-singlet Cooper pair crosses the interface from SC to FM, the two electrons enter into different spin bands in the FM and the pair wavefunction acquires a center-of-mass monentum leading to an oscillatory, decaying superconducting gap in the FM[50]. Previous works have proven that long-ranged proximity effect may be induced in FMs in mesoscopic SC-FM hybrid structures when the contact region shows inhomogeneous magnetic polarization and small resistance [51]. Whether the long range proximity effect survives in 1D nanowires remained to be seen. To explore this possibility, experiments were conducted on ferromagnetic Co and Ni nanowires.

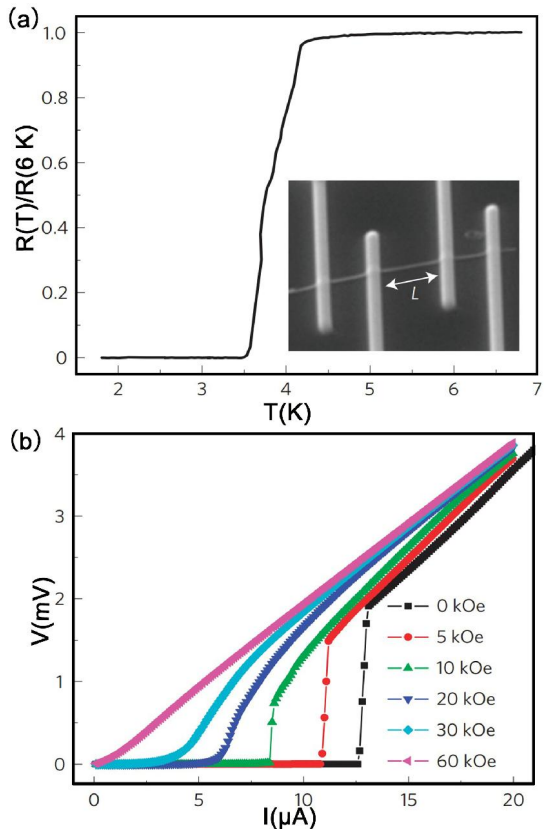

**Fig. 14 (a)** Zero resistance is found below 3.5 K. The inset is a SEM of the Co nanowire contacted by four FIB-deposited superconducting W electrodes. **(b)** Voltage vs. current curves of the Co nanowire measured at different perpendicular magnetic fields at 1.8 K. The lengths of the nanowires (L) in this article are defined to be the distance between the inner edges of the voltage electrodes. The resistance at 6 K is 193 Ω and the resistivity (ρ) of the wire, assuming an oxide shell of 2 nm, is 32 μΩcm. This figure is from Ref. [52].

The crystalline Co and Ni nanowires used in our experiments were fabricated by the template based electrodeposition technique referred to for the Bi and Au nanowires. Individual nanowires were contacted by four superconducting W electrodes using FIB assisted deposition for conventional four-probe measurements (inset of Fig. 14 (a)). The length L of the nanowire is defined as the distance between the inner two (voltage) electrodes. Short Co nanowire (600 nm long, 40 nm in diameter) exhibit typical supercondcuting properties (see Fig. 14) with an onset $T_C$ = 4.2 K and zero resistance below $T_C$ = 3.5 K. The induced superconductivity is also suppressed by magnetic fields. This indicates that the spatial extent of the proximity effect induced in Co nanowire is at least 300 nm.

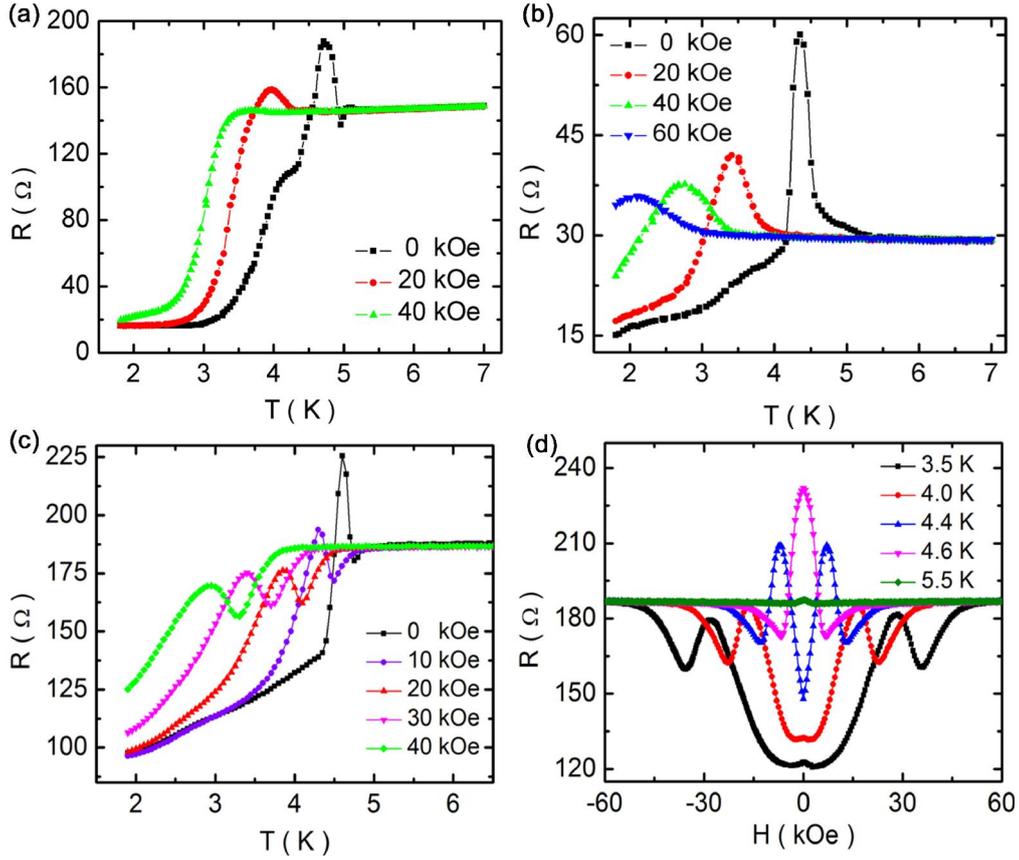

**Fig. 15 (a)** $R$ vs. $T$ curves at different fields for an individual 40 nm Co nanowire with L= 1.5 μm. ρ at 6 K is 10 μΩcm. **(b)** Resistance as a function of temperature at different fields for an individual 80 nm Co nanowire with L=1.5 μm. ρ at 6 K is 9 μΩcm. **(c)** Resistance as a function of temperature at different fields for an individual 60 nm Ni nanowire with L= 3 μm. ρ at 6 K is 15.3 μΩcm. **(d)** Resistance vs. magnetic field of a 60 nm Ni nanowire, with L= 3 μm. This figure is from Ref. [52].

Resistance vs. temperature plots for 1.5 μm Co nanowire of 40 nm (Fig. 15(a) and 80 nm in diameter(Fig. 15(b))) show large resistance peaks (25% of the normal-state resistance for the 40 nm wire and ∼ 100% for the 80 nm wire) just above the temperature at the superconducting resistance drop. Measurements in both warming and cooling scans show an absence of hysteretic behaviour. The residual resistance of the 40 nm (80 nm) wire is 11% (50%) of the normal state value even at 2 K, indicating the coexistence of superconductivity and ferromagnetism. With an increasing magnetic field, the peak moves to lower temperature and the magnitude is suppressed.

Other observations of resistance peaks in superconducting granular metal thin films with homogeneous disorder (R□ = 6.45 kΩ) are attributed to electron localization [53]. Nevertheless, for our individual 40 nm Co nanowire with L=1.5 μm, ρ is 10 μΩcm (6 K), far less than their value 6.45 kΩ. If the ferromagnetic nanowires with normal metal or superconduting nanowires, the resistance peak is not seen[46; 47]. Spin accumulation [54; 55] is another possible mechanism for the origin for the resistance peak observed as seen in mesoscopic Fe-In junctions [56]. In that case, the absolute peak value was small ($10^{-8}\Omega$), and the relative change(ΔR/R) is just 0.05%. However, for our samples, the spin-accumulation mechanism predicts a substantially smaller resistance peak in the 1.5 μm nanowire (4%) than that observed in our experiments (25% ∼ 100%). Furthermore, the spin-accumulation model assumes that the induced superconductivity is singlet, which seems inconsistent with the long-ranged nature of the proximity effect. Therefore, it seems spin accumulation cannot account for both the large resistance peak and the long-ranged proximity effect simultaneously. Resistance peaks in mesoscopic Al wires [57] are explained simply in terms of a nonequilibrium charge-imbalance model. The peaks induced by charge-imbalance model were found to be very sensitive to the applied magnetic field, suppressed by a tiny field of approximately 10 Oe. In contrast, the resistance peaks observed in our Co and Ni nanowires are distinct from the peak effect above as they are robust and persist even in the presence of a large (several Tesla) magnetic field. As a result none of the above theories fit our results.

Similar critical peak around $T_C$ and incomplete superconducting resistance drop below $T_C$ were also observed in ferromagnetic Ni (3 μm long) nanowires with a diameter of 60 nm contacted by the same superconducting electrodes (Fig. 15(c)). The low temperature residual resistance is 52% of the normal state resistance, which suggests that the spatial extent of the proximity effect is again several hundred nanometres. In addition in the magnetoresistance behavior of the Ni nanowire (Fig. 14 (d)) AMR effect was observed in small fields in the nominally 'superconducting' low resistance state. The presence of AMR confirms that the wire do have ferromagnetic order.

The long range proximity effect revealed in crystalline ferromagnetic nanowires may offer the possibility of combination of the zero-resistance supercurrent and the spin alignment in same sample, which would pave the way for spin-polarized supercurrent for new spintronics [58].

───────────────────────────────────

## 4 Interplay between nanostructured TIs and superconducting electrodes

The combination of SCs and TIs holds interesting prospect for both fundamental physics and applications/new technology. The interplay between the topological order and symmetry breaking that appears in the ordered phases of SCs may lead to many proposals of novel quantum phenomena such as non-Abelian Majorana fermions [59; 60]. There have been a great deal of experimental effort has been expended on SC-TI hybrid structures in recent years [61; 62; 63; 64]. Here we mainly focus on the interplay between nanostructured TIs and superconducting electrodes. What will happen if we use superconducting electrodes to contact topological insulator film or nanobelt? Can superconductivity seep into TI film and nanobelt?

Here we mainly focus on the interplay between nanostructured TIs and superconducting electrodes. The question we seek to answer is whether superconducting order from the electrodes can survive in the spin polarized surface states of a TI film or nanobelt.

We use three kinds of superconducting electrodes to contact MBE-grown $Bi_2Se_3$ thin films. Bulk Indium (In), mesoscopic Aluminum (Al) and W were attached onto the surface of the $Bi_2Se_3$ film by mechanical pressure (In), electron-beam lithography and evaporation (Al) and FIB deposition (W). The distances between two electrodes are 1 mm (In) and 1 μm (Al and W). The thicknesses of three films are 5 nm (In), 200 nm (Al), 200 nm (W).

The R vs. T curves all show abrupt and significant upturns in resistance in the three experimental structures (Fig. 16). The upturn temperatures are 3.29 K (In), 0.95 K (Al), 3.5 K (W), respectivly. These resistance upturns decrease rapidly with increasing magnetic field. For $Bi_2Se_3$ film contacted by two superconducting In dots (Fig. 16(b)), when magnetic field reaches 200 Oe, the upturn behavior is suppressed absolutely. The upturn in resistance of $Bi_2Se_3$ film corresponds closely to the onset temperature ($T_C$s) and magnetic field ($H_C$s) of the superconducting electrodes. However the $T_C$s and $H_C$s extracted from the upturn in resistance are slightly lower than the $T_C$s and $H_C$s of bulk In, Al and W determined independently. We attribute this reduction in critical transition temperatures and fields to the $Bi_2Se_3$ film greatly weakening the superconductivity of mesoscopic superconducting electrodes. For example, the $T_C$ of bulk In, bulk Al and W stripe are 3.4 K, 1.2 K and 4 K, respectivily corresponding to the onset Tc of resistance upturns of 3.29 K (In), 0.95 K (Al), 3.5 K (W).

One possible interpretation of this result is that the superconducting electrodes are accessing the TI surface states. These experimental results stem from the interplay between the Cooper pairs of the electrodes having anti-parallel spins and the spin-polarized current of the surface states in $Bi_2Se_3$ film requiring all electrons to have parallel spins. In this transport measurement configuration, the spin polarization of the TI surface state is decided by the current direction. Electron in each spin–singlet Cooper pair is not compatible with this spin-polarization upon arriving at the TI interface. Spin flip processes must take place at the interface when the spin-polarized electrons flow from TI to the superconducting electrode or when the Cooper pairs leak from the current source electrode to TI. This process produces an abrupt resistance upturn just as the electrodes turn superconducting. The spin-polarized current in turn strongly weakens the superconductivity of the electrodes.

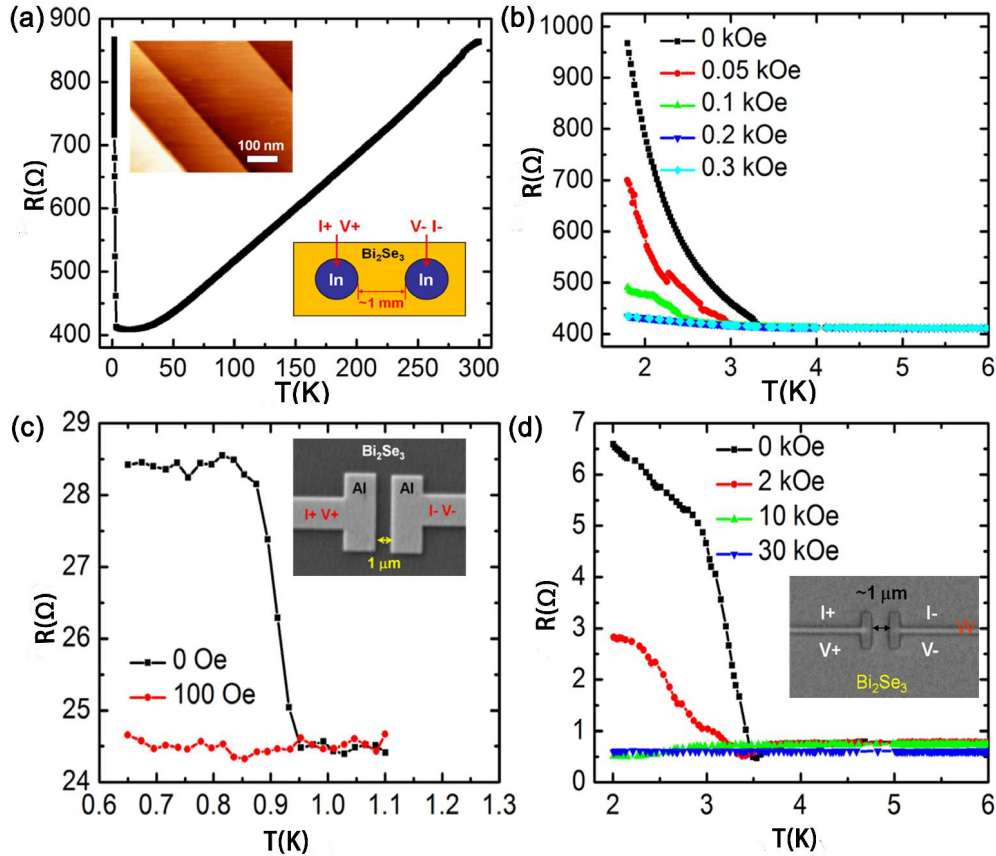

**Fig. 16 (a)** $R$ vs. $T$ behavior of the 5-nm-thick $Bi_2Se_3$ film contacted by two superconducting In dots. The right inset is the measurement structure. **(b)** $R$ vs. $T$ behaviors of 5-nm-thick $Bi_2Se_3$ film contacted by two superconducting In dots at different perpendicular fields. **(c)** Transport behaviors of 200-nm-thick $Bi_2Se_3$ films contacted by superconducting Al electrodes. An applied magnetic field of 100 Oe suppresses the enhancement. The inset is a SEM image of the Al contacts on the surface of the $Bi_2Se_3$ film. **(d)** Transport behaviors of a 200-nm-thick $Bi_2Se_3$ film contacted by superconducting W electrodes. $R$ vs. $T$ scans under different magnetic fields. The inset is a SEM image of the W contacts on the surface of the $Bi_2Se_3$ film. This figure is from Ref.[65].

We also find the superconducting proximity effect in mesoscopic $Bi_2Se_3$ nanoribbons by contacting them using superconducting W electrodes [66]. The $Bi_2Se_3$ nanoribbons were made using gold catalyzed vapor–liquid–solid mechanism in a horizontal tube furnace [10].

Temperature dependence of the two-probe zero-bias differential resistance of device A (width =600 nm, thickness = 60 nm, length = 1.08μm) is shown in Fig. 17(a). The onset of the proximity effect is at T ~ 4.7 K when the W contacts become superconducting, eventually transitioning to a zero-resistance superconducting state at T = 2 K. Figure 17(b) shows *I* vs. *V* characteristics at different temperatures. The critical current $I_C$ is 1.1 μA at 500 mK. $I_C$ decreases with increasing temperature until 2 K. In the d*I*/d*V* vs. *V* curve (see Fig. 17(c)), we can see subharmonic gap structure due to multiple Andreev reflections at V = 2Δ/ne (Δ is the superconducting gap of W electrodes, n is an integer). According to the relationship between the position of differential conductance V and 1/n, we calculate $T_C$ (4.41 K) from BCS relation (Δ = 1.73 $k_B T_C$), which is closed to $T_C$ (4.7 K) obtained from the temperature dependent zero-bias differential resistance.

Our SC-TI nanoribbon device configuration provides a viable route for long range proximity effect in TI and paves a way to explore Majorana fermions.

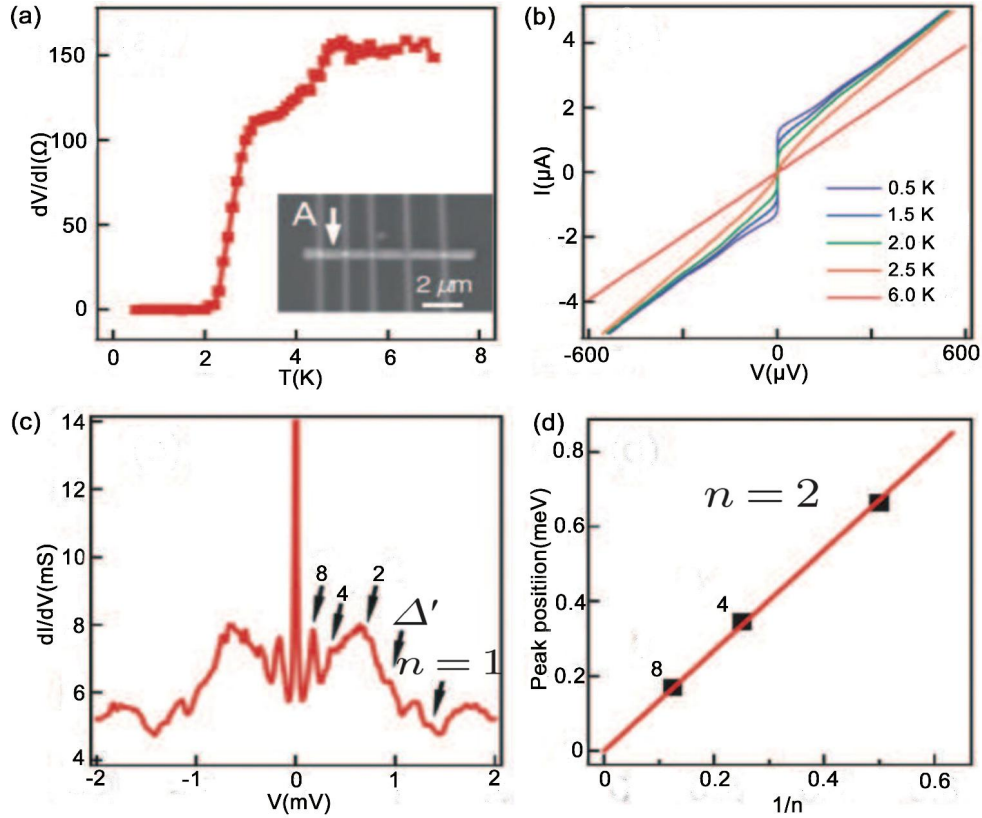

**Fig. 17 (a)** dV/dI vs. temperature for device A at H = 0. Inset shows an SEM image of the device, with the arrow indicating the measured channel with edge-to-edge length of 1.08 μm between two W electrodes. **(b)** I-V characteristics at various temperatures, measured using the same contacts as in (a). **(c)** dI/dV vs. V in device A at T = 500 mK and in zero magnetic field. The arrows identify a consistent subharmonic series of conductance anomalies corresponding to sub harmonic gap structure (2Δ/ne with n = 2, 4, 8). **(d)** Position of differential conductance anomalies as a function of the index 1/n. This figure is from Ref. [66]

## 5 Summary

Remarkable progresses in TIs and SCs have been made in the past few years. For example, in TI $Bi_2Se_3$ thin films, EEI, WAL and AMR property have been found. The observation of period magneto-resistance oscillations in superconducting Pb nanobridge and the observed superconductivity in Bi nanowires reveals exotic nanoscale properties. Furthermore, interesting superconducting proximity effect in non-superconducting nanowires, as well as TI films and nanoribbons have been systematically studied.

## Acknowledgements

We want to thank the folks at MCL and nanofab of the Pennsylvania State University for their help in the fabrication of the nanowire samples. Also we want to thank Qi-Kun Xue, Ke He, Cui-Zu Chang, Nitin Samarth, Jainendra Jain, Mao-Hai Xie, Mingliang Tian, Chuntai Shi, T. E. Mallouk, Duming Zhang, Ashley M. DaSilva, Handong Li, Xu-Cun Ma, Yun Qi, ShuaihuaJi, Yingshuang Fu, Bangzhi Liu, Joon Sue Lee, Li Lu, Ai-Zi Jin, Chang-Zhi Gu, Qi Zhang, Xi Chen, Jin-Feng Jia, X.C.Xie, Shunqing Shen, Haizhou Lu, Lin He, Nitesh Kumar-a necessarily incomplete list. This work was financially supported by the National Basic Research Program of China (Grant Nos. 2013CB934600 & 2012CB921300), the National Natural Science Foundation of China (Nos. 11222434 & 11174007), the Pennsylvania State University Materials Research Science and Engineering Center under National Science Foundation Grant No. DMR-0820404, and China Postdoctoral Science Foundation (No. 2011M500180 & No. 2012T50012).

## References

[1] M.Z. Hasan, and C.L. Kane, Physics 82 (2010) 3045.

[2] S. Murakami, N. Nagaosa, and S.-C. Zhang, Physical Review Letters 93 (2004) 156804.

[3] D. Hsieh, D. Qian, L. Wray, Y. Xia, Y.S. Hor, R.J. Cava, and M.Z. Hasan, Nature 452 (2008) 970.

[4] Y. Xia, D. Qian, D. Hsieh, L. Wray, A. Pal, H. Lin, A. Bansil, D. Grauer, Y.S. Hor, R.J. Cava, and M.Z. Hasan, Nature Physics 5 (2009) 398.

[5] P. Roushan, J. Seo, C.V. Parker, Y.S. Hor, D. Hsieh, D. Qian, A. Richardella, M.Z. Hasan, R.J. Cava, and A. Yazdani, Nature 460 (2009) 1106.

[6] T. Zhang, P. Cheng, X. Chen, J.-F. Jia, X. Ma, K. He, L. Wang, H. Zhang, X. Dai, Z. Fang, X. Xie, and Q.-K. Xue, Physical Review Letters 103 (2009)266803.

[7] J. Chen, H.J. Qin, F. Yang, J. Liu, T. Guan, F.M. Qu, G.H. Zhang, J.R. Shi, X.C. Xie, C.L. Yang, K.H. Wu, Y.Q. Li, and L. Lu, Physical Review Letters 105 (2010) 176602.

[8] J.G. Checkelsky, Y.S. Hor, M.H. Liu, D.X. Qu, R.J. Cava, and N.P. Ong, Physical Review Letters 103 (2009)246601.

[9] D.X. Qu, Y.S. Hor, J. Xiong, R.J. Cava, and N.P. Ong, Science 329 (2010) 821.

[10] H.L. Peng, K.J. Lai, D.S. Kong, S. Meister, Y.L. Chen, X.L. Qi, S.C. Zhang, Z.X. Shen, and Y. Cui, Nature Materials 9 (2010) 225.

[11] C.Z. Chang, J.S. Zhang, X. Feng, J. Shen, Z.C. Zhang, M.H. Guo, K. Li, Y.B. Ou, P. Wei, L.L. Wang, Z.Q. Ji, Y. Feng, S.H. Ji, X. Chen, J.F. Jia, X. Dai, Z. Fang, S.C. Zhang, K. He, Y.Y. Wang, L. Lu, X.C. Ma, and Q.K. Xue, Science 340 (2013) 167.

[12] Y. Liu, Z. Ma, Y.-F. Zhao, M. Singh, and J. Wang, Chinese Physics B 22 (2013) 067302.

[13] M. Konig, S. Wiedmann, C. Brune, A. Roth, H. Buhmann, L.W. Molenkamp, X.L. Qi, and S.C. Zhang, Science 318 (2007) 766.

[14] H.J. Zhang, C.X. Liu, X.L. Qi, X. Dai, Z. Fang, and S.C. Zhang, Nature Physics 5 (2009) 438.

[15] H.Z. Lu, J.R. Shi, and S.Q. Shen, Physical Review Letters 107 (2011)076801.

[16] M.H. Liu, J.S. Zhang, C.Z. Chang, Z.C. Zhang, X. Feng, K. Li, K. He, L.L. Wang, X. Chen, X. Dai, Z. Fang, Q.K. Xue, X.C. Ma, and Y.Y. Wang, Physical Review Letters 108 (2012)036805.

[17] J. Chen, H.J. Qin, F. Yang, J. Liu, T. Guan, F.M. Qu, G.H. Zhang, J.R. Shi, X.C. Xie, C.L. Yang, K.H. Wu, Y.Q. Li, and L. Lu, Physical Review Letters 105 (2010) 176602.

[18] M. Liu, C.-Z. Chang, Z. Zhang, Y. Zhang, W. Ruan, K. He, L.-l. Wang, X. Chen, J.-F. Jia, S.-C. Zhang, Q.-K. Xue, X. Ma, and Y. Wang, Physical Review B 83 (2011) 165440.

[19] H.-T. He, G. Wang, T. Zhang, I.-K. Sou, G.K.L. Wong, J.-N. Wang, H.-Z. Lu, S.-Q. Shen, and F.-C. Zhang, Physical Review Letters 106 (2011) 166805.

[20] M. Tian, W. Ning, Z. Qu, H. Du, J. Wang, and Y. Zhang, Scientific reports 3 (2013) 1212.

[21] J. Wang, A.M. DaSilva, C.-Z. Chang, K. He, J.K. Jain, N. Samarth, X.-C. Ma, Q.-K. Xue, and M.H.W. Chan, Physical Review B 83 (2011)245438.

[22] S. Hikami, A.I. Larkin, and Y. Nagaoka, Progress of Theoretical Physics 63 (1980) 707-710.

[23] S. Maekawa, and H. Fukuyama, Journal of the Physical Society of Japan 50 (1981) 2516.

[24] P.A. Lee, and T.V. Ramakrishnan, Physical Review B 26 (1982) 4009.

[25] Z. Zeng, T.A. Morgan, D. Fan, C. Li, Y. Hirono, X. Hu, Y. Zhao, J.S. Lee, J. Wang, Z.M. Wang, S. Yu, M.E. Hawkridge, M. Benamara, and G.J. Salamo, AIP Advances 3 (2013) 072112.

[26] J. Wang, H. Li, C. Chang, K. He, J.S. Lee, H. Lu, Y. Sun, X. Ma, N. Samarth, S. Shen, Q. Xue, M. Xie, and M.H.W. Chan, Nano Research 5 (2012) 739.

[27] J. Wang, X.-C. Ma, Y. Qi, Y.-S. Fu, S.-H. Ji, L. Lu, J.-F. Jia, and Q.-K. Xue, Applied Physics Letters 90 (2007) 113109.

[28] J. Wang, X.C. Ma, Y. Qi, Y.S. Fu, S.H. Ji, L. Lu, X.C. Xie, J.F. Jia, X. Chen, and Q.K. Xue, Nanotechnology 19 (2008) 475708.

[29] J. Wang, X.-C. Ma, L. Lu, A.-Z. Jin, C.-Z. Gu, X.C. Xie, J.-F. Jia, X. Chen, and Q.-K. Xue, Applied Physics Letters 92 (2008) 233119.

[30] J. Wang, X. Ma, S. Ji, Y. Qi, Y. Fu, A. Jin, L. Lu, C. Gu, X.C. Xie, M. Tian, J. Jia, and Q. Xue, Nano Research 2 (2009) 671.

[31] W.A. Little, and R.D. Parks, Observation of Physical Review Letters 9 (1962) 9.

[32] M.L. Tian, J. Wang, Q. Zhang, N. Kumar, T.E. Mallouk, and M.H.W. Chan, Nano letters 9 (2009) 3196.

[33] M.L. Tian, J.G. Wang, N. Kumar, T.H. Han, Y. Kobayashi, Y. Liu, T.E. Mallouk, and M.H.W. Chan, Nano letters 6 (2006) 2773-2780.

[34] Y. Eckstein, and J.B. Ketterson, Physical Review 137 (1965) A1777.

[35] P. De Gennes, Reviews of Modern Physics 36 (1964) 225.

[36] J. Wang, Y. Sun, M. Tian, B. Liu, M. Singh, and M.H.W. Chan, Physical Review B 86 (2012)035439.

[37] M. Tian, N. Kumar, S.Y. Xu, J.G. Wang, J.S. Kurtz, and M.H.W. Chan, Physical Review Letters 95 (2005)076802.

[38] M.L. Tian, N. Kumar, J.G. Wang, S.Y. Xu, and M.H.W. Chan, Physical Review B 74 (2006)014515.

[39] M. Singh, J. Wang, M.L. Tian, T.E. Mallouk, and M.H.W. Chan, Physical Review B 83 (2011)220506.

[40] Y. Chen, Y.H. Lin, S.D. Snyder, and A.M. Goldman, Physical Review B 83 (2011)054505.

[41] Y. Chen, S.D. Snyder, and A.M. Goldman, Physical Review Letters 103 (2009)127002.

[42] M. Singh, J. Wang, M.L. Tian, Q. Zhang, A. Pereira, N. Kumar, T.E. Mallouk, and M.H.W. Chan, Chem. Mat. 21 (2009) 5557.

[43] H.C. Fu, A. Seidel, J. Clarke, and D.H. Lee, Physical Review Letters 96 (2006)157005.

[44] A.O. Caldeira, and A.J. Leggett, Physical Review Letters 46 (1981) 211.

[45] M.L. Tian, J.U. Wang, J. Kurtz, T.E. Mallouk, and M.H.W. Chan, Nano letters 3 (2003) 919.

[46] Y. Sun, J. Wang, W. Zhao, M. Tian, M. Singh, and M.H.W. Chan, Scientific reports 3 (2013)2307.

[47] J. Wang, C. Shi, M. Tian, Q. Zhang, N. Kumar, J. Jain, T. Mallouk, and M. Chan, Physical Review Letters 102 (2009)247003.

[48] L. He, and J. Wang, Nanotechnology 22 (2011) 445704.

[49] A.I. Buzdin, Reviews of Modern Physics 77 (2005) 935.

[50] E.A. Demler, G.B. Arnold, and M.R. Beasley, Physical Review B 55 (1997) 15174.

[51] M. Giroud, H. Courtois, K. Hasselbach, D. Mailly, and B. Pannetier, Physical Review B 58 (1998) 11872.

[52] J. Wang, M. Singh, M. Tian, N. Kumar, B.Z. Liu, C. Shi, J.K. Jain, N. Samarth, T.E. Mallouk, and M.H.W. Chan, Nature Physics 6 (2010)

389.

[53] H.M. Jaeger, D.B. Haviland, B.G. Orr, and A.M. Goldman, Physical Review B 40 (1989) 182.

[54] F.J. Jedema, B.J. van Wees, B.H. Hoving, A.T. Filip, and T.M. Klapwijk, Physical Review B 60 (1999) 16549.

[55] V.I. Fal'ko, A.F. Volkov, and C. Lambert, Physical Review B 60 (1999) 15394.

[56] Y.N. Chiang, O.G. Shevchenko, and R.N. Kolenov, Low Temperature Physics 33 (2007) 314.

[57] P. Santhanam, C. Chi, S. Wind, M. Brady, and J. Bucchignano, Physical Review Letters 66 (1991) 2254.

[58] M. Eschrig, Physics Today 64 (2011) 43.

[59] L. Fu, and C.L. Kane, Physical Review Letters 100 (2008) 096407.

[60] J. Linder, Y. Tanaka, T. Yokoyama, A. Sudbø, and N. Nagaosa, Physical Review Letters 104 (2010) 067001.

[61] B. Sacepe, J.B. Oostinga, J. Li, A. Ubaldini, N.J.G. Couto, E. Giannini, and A.F. Morpurgo, Nature communications 2 (2011) 575.

[62] M. Veldhorst, M. Snelder, M. Hoek, T. Gang, V.K. Guduru, X.L. Wang, U. Zeitler, W.G. van der Wiel, A.A. Golubov, H. Hilgenkamp, and A. Brinkman, Nature Materials 12 (2013) 171.

[63] M.X. Wang, C.H. Liu, J.P. Xu, F. Yang, L. Miao, M.Y. Yao, C.L. Gao, C.Y. Shen, X.C. Ma, X. Chen, Z.A. Xu, Y. Liu, S.C. Zhang, D. Qian, J.F. Jia, and Q.K. Xue, Science 336 (2012) 52.

[64] Y.X. Ou, M. Singh, and J. Wang, Science China-Physics Mechanics & Astronomy 55 (2012) 2226.

[65] J. Wang, M. Singh, M. Tian, N. Kumar, B. Liu, C. Shi, J.K. Jain, N. Samarth, T.E. Mallouk, and M.H.W. Chan, Nature Physics 6 (2010) 389.

[66] D. Zhang, J. Wang, A.M. DaSilva, J.S. Lee, H.R. Gutierrez, M.H.W. Chan, J. Jain, and N. Samarth, Physical Review B 84 (2011) 165120.